\documentclass{webofc}
\usepackage[varg]{txfonts}   
\def\eps{\varepsilon}
\begin{document}
\title{A tale of two Regge limits}

\author{\firstname{~} \lastname{Vittorio Del Duca}\inst{1,2}\fnsep\thanks{\email{delducav@itp.phys.ethz.ch}}    }

\institute{Institute for Theoretical Physics, ETH Z\"urich, 8093 Z\"urich, Switzerland
\and
           INFN, Laboratori Nazionali di Frascati, Italy
          }

\abstract{%
  In light of the strong advances in understanding the mathematical structure of scattering amplitudes,
 we discuss the Regge limit of QCD and of the ${\cal N}=4$ Super Yang-Mills theory.
}
\maketitle

\section{Regge limit in QCD}
\label{reggeqcd}

In the Regge limit, in which the squared centre-of-mass energy $s$ is much 
larger than the momentum transfer $|t|$, $s\gg |t|$, any scattering process is dominated by the exchange in the $t$ channel
of the highest-spin particle. In the case of QCD or ${\cal N}=4$ Super Yang-Mills (SYM), that entails the exchange of a gluon.
The Regge limit of QCD is described by the Balitsky-Fadin-Kuraev-Lipatov (BFKL) theory, 
which models strong-interaction processes with two large and disparate scales, by resumming the radiative corrections to parton-parton scattering.
This is achieved at leading logarithmic (LL) accuracy, in $\ln(s/|t|)$, through the BFKL 
equation~\cite{Fadin:1975cb,Kuraev:1976ge,Kuraev:1977fs,Balitsky:1978ic}, {\it i.e.} an integral equation for the evolution of the 
$t$-channel gluon propagator in transverse-momentum space and Mellin moment space. The building blocks of the kernel of the BFKL equation are
a real correction -- the emission of a gluon along the ladder, {\it a.k.a.} the central emission vertex -- and a virtual correction -- the 
one-loop Regge trajectory~\cite{Lipatov:1976zz}. The BFKL equation is then obtained by iterating the one-loop corrections to all orders in $\alpha_S$,
at LL accuracy in $\ln(s/|t|)$, {\it i.e.} by resumming the terms of ${\cal O}(\alpha_S^n \ln^n(s/|t|))$. 
Underpinning the BFKL theory is the ladder structure in the $t$ channel of the scattering amplitudes in the Regge limit. For example,
in the case of gluon-gluon scattering in QCD, the gluon ladder is described by the 
${\mathbf 8} \otimes {\mathbf 8}$ colour representation, which is decomposed as ${\mathbf 8} \otimes {\mathbf 8} = \{{\mathbf 1} \oplus {\mathbf 8}_s \oplus \mathbf{27}\} \oplus [{\mathbf 8}_a \oplus \mathbf{10} \oplus \mathbf{\overline{10}}]$,
where the term in curly brackets in the direct sum is even in colour space under $s\leftrightarrow u$ exchange,
while the term in square brackets is odd. However, to LL accuracy in $\ln(s/|t|)$, the parton-parton scattering amplitudes
are real and only the antisymmetric octet ${\mathbf 8}_a$ contributes. That implies the gluon Reggeisation at LL accuracy~\cite{Lipatov:1976zz}.
For phenomenological purposes,
the BFKL equation needs then be endowed with process-dependent impact factors, the simplest case of which are the quark or gluon impact
factors~\cite{Kuraev:1976ge,DelDuca:1995zy}, by which one can obtain, for example, the dijet cross section at large rapidities~\cite{Mueller:1986ey}.

The computation of the next-to-leading-logarithmic (NLL) corrections~\cite{Fadin:1998py,Ciafaloni:1998gs,Kotikov:2000pm,Kotikov:2002ab}
to the BFKL equation, {\it i.e.} of the terms of ${\cal O}(\alpha_S^n \ln^{n-1}(s/|t|))$, requires computing the radiative corrections to the kernel of
the BFKL equation, {\it i.e.} the real corrections to the central emission vertex 
-- the emission of two gluons or of a quark-antiquark pair along the ladder~\cite{Fadin:1989kf,DelDuca:1995ki,Fadin:1996nw,DelDuca:1996nom} 
-- and the one-loop corrections to the central emission vertex~\cite{Fadin:1993wh,Fadin:1994fj,Fadin:1996yv,DelDuca:1998cx,Bern:1998sc},
as well as the two-loop Regge trajectory~\cite{Fadin:1995xg,Fadin:1996tb,Fadin:1995km,Blumlein:1998ib,DelDuca:2001gu}.
Underpinning the BFKL theory at NLL accuracy is the fact that although the parton-parton scattering amplitudes develop an imaginary part,
the BFKL equation is computed through the real part of the amplitudes, to which only the antisymmetric octet ${\mathbf 8}_a$ contributes, 
which implies the gluon Reggeisation at NLL accuracy~\cite{Fadin:2006bj,Fadin:2015zea}.

At LL accuracy, the kernel of the BFKL equation is conformally invariant, and thus the leading-order eigenfunctions of the 
BFKL equation are fixed by conformal symmetry~\cite{Lipatov:1985uk}. The NLL corrections~\cite{Fadin:1998py,Ciafaloni:1998gs,Kotikov:2000pm,Kotikov:2002ab} to the BFKL equation were computed
by acting with the next-to-leading-order (NLO) kernel of the equation on the leading-order eigenfunctions. This procedure is not consistent, and it was already clear to Fadin and Lipatov~\cite{Fadin:1998py} that the terms which make the procedure inconsistent are related to the running of the coupling.
The consistent NLO eigenfunctions were constructed by Chirilli and Kovchegov~\cite{Chirilli:2013kca,Chirilli:2014dcb}, who found indeed that the 
additional pieces which occur at NLO are proportional to the beta function.
However, if the scale of the strong coupling is chosen to be the geometric mean of the transverse momenta at the ends of the ladder, 
then one can use the leading-order eigenfunctions instead of the NLO ones~\cite{DelDuca:2017peo}.

\subsection{Regge limit and infrared factorisation}

The backbone of the BFKL equation are the quark and gluon scattering amplitudes, which are dominated by the
exchange of a gluon in the $t$ channel.
Because loop-level scattering amplitudes of massless partons are infrared divergent, and so are the impact factors, the Regge trajectory,
and the central emission vertex, in which they factorise in the Regge limit, the study of the scattering amplitudes in
the Regge limit has benefited from a cross breeding with infrared 
factorisation~\cite{Sotiropoulos:1993rd,Korchemsky:1993hr,Korchemskaya:1994qp,Korchemskaya:1996je,Bret:2011xm,DelDuca:2011ae,Caron-Huot:2013fea,DelDuca:2013ara,DelDuca:2014cya,Caron-Huot:2017fxr,Caron-Huot:2017zfo},
where the infrared structure of scattering amplitudes for massless partons is known up to three 
loops~\cite{Aybat:2006wq,Aybat:2006mz,Becher:2009cu,Gardi:2009qi,Almelid:2015jia,Almelid:2017qju}.

For example, the one-loop impact factors~\cite{Fadin:1992zt,Fadin:1993wh,Fadin:1993qb,DelDuca:1998kx,Bern:1998sc} and the 
two-loop Regge trajectory have poles in the dimensional regulator $\eps$ in $d=4-2\eps$ dimensions,
which can be understood, for the one-loop impact factors, in terms of the one-loop cusp anomalous dimension and the one-loop 
quark and gluon collinear anomalous dimensions~\cite{DelDuca:2013ara,DelDuca:2014cya}, and for the two-loop Regge trajectory 
in terms of the two-loop cusp anomalous dimension~\cite{Korchemskaya:1996je}.

Further, we know that the picture of the Regge-pole factorisation of the amplitudes, based on the gluon Reggeisation, breaks down 
at two loops by $N_c$-subleading terms~\cite{DelDuca:2001gu}.
The violation can be explained through infrared factorisation by showing that the real part of the amplitudes 
becomes non-diagonal in the $t$-channel-exchange basis~\cite{Bret:2011xm,DelDuca:2011ae}.
Accordingly, it can be predicted how the violation propagates to higher loops, and the three-loop prediction 
for the violation~\cite{DelDuca:2013ara,DelDuca:2014cya}, which has NNLL accuracy, has been confirmed by the explicit computation of the 
three-loop four-point function of full ${\cal N}=4$ SYM~\cite{Henn:2016jdu}. 
In the Regge factorisation picture, that violation is due to a Regge cut and to the exchange of three Reggeised gluons~\cite{Fadin:2016wso,Caron-Huot:2017fxr}. Finally, at three loops and at NNLL accuracy, also the $\mathbf{10} \oplus \mathbf{\overline{10}}$ colour representation contributes to the exchange of three Reggeised gluons~\cite{Caron-Huot:2017fxr}.

The imaginary part of the amplitude is even under $s\leftrightarrow u$ exchange. For gluon-gluon scattering,
the ${\mathbf 8}_s$ colour representation contributes only at one loop to it through a Regge pole {\it i.e.} through the exchange of a Reggeised gluon,
while more in general at NLL accuracy the ${\mathbf 1}$ and the $\mathbf{27}$ representations contribute through a Regge cut and the exchange of two Reggeised gluons~\cite{Caron-Huot:2017zfo}.
The imaginary part of the amplitude contributes then to the squared amplitude, and thus to the BFKL ladder, at NNLL accuracy.

Thus, although a study of the BFKL ladder and firstly of its building blocks at NNLL accuracy is yet to be undertaken -- the three-loop Regge trajectory
in a particular scheme~\cite{Caron-Huot:2017fxr} and the emission of three partons along the gluon ladder~\cite{DelDuca:1999iql} being the only ones computed so far -- we already have a clear picture of which contributions occur at NNLL accuracy.

While we have a precise knowledge of how the infrared poles occur in the loop corrections
to the impact factors and to the Regge trajectory, the finite parts of those corrections are treated as free parameters,
which neither infrared nor Regge factorisation can constrain. However, there is a relation~\cite{DelDuca:2017pmn}
between the ${\cal O}(\eps)$ terms of the one-loop gluon impact factor~\cite{Bern:1998sc} and the ${\cal O}(\eps^0)$ terms of 
the two-loop Regge trajectory~\cite{Fadin:1995xg,Fadin:1996tb,Fadin:1995km,Blumlein:1998ib,DelDuca:2001gu},
which hints at an exponentiation of some finite parts of the amplitude in the Regge limit, and thus at more structure in infrared factorisation 
in the Regge limit than we currently know.

\section{Regge limit in ${\cal N}=4$ SYM}

In the last few years, it has been realised that the Regge limit of QCD and of ${\cal N}=4$ SYM, and thus the BFKL equation, 
are endowed with a rich mathematical structure.

In the planar limit of ${\cal N}=4$ SYM, scattering amplitudes exhibit a dual conformal 
symmetry~\cite{Drummond:2006rz,Alday:2007hr,Drummond:2007aua,Brandhuber:2007yx}.
The dual conformal invariance is broken by infrared divergences~\cite{Drummond:2007cf,Drummond:2007au}, but it is possible to construct finite, dual conformally invariant ratios in which all infrared divergences cancel. As a consequence, the analytic structure of scattering amplitudes in planar ${\cal N}=4$
SYM is highly constrained. In particular, the four and five-point amplitudes are fixed to all loop orders by symmetries in terms of the one-loop amplitudes and the cusp anomalous dimension~\cite{Drummond:2007au,Anastasiou:2003kj,Bern:2005iz}. Non-trivial corrections occur for amplitudes with at least six legs~\cite{Drummond:2007au,Bern:2008ap,Drummond:2008aq,DelDuca:2009au,DelDuca:2010zg}.
The ordinary and dual conformal symmetries are also at the heart of a duality between  scattering amplitudes 
and Wilson loops computed along a light-like polygonal 
contour~\cite{Alday:2007hr,Drummond:2007aua,Brandhuber:2007yx,Drummond:2007au,Drummond:2007cf,Drummond:2008aq}.

It has been observed that the four-point amplitude is Regge exact~\cite{Drummond:2007aua,Naculich:2007ub}, 
{\it i.e.} it is not modified by the Regge limit. The Regge exactness extends to the 
five-point amplitude~\cite{Brower:2008nm,Bartels:2008ce}, if computed in the multi-Regge kinematics (MRK)~\cite{Lipatov:1976zz}, 
which features the central-emission vertex; and to the six-point amplitude~\cite{DelDuca:2009au}, 
if computed in the quasi-multi-Regge kinematics~\cite{Fadin:1989kf},
which features the emission of two gluons along the ladder~\cite{Fadin:1989kf,DelDuca:1995ki}. 
In fact, Regge exactness extends in general to the $n$-point MHV amplitudes,
if computed in the quasi-multi-Regge kinematics of the emission of $(n-4)$ gluons along the ladder~\cite{DelDuca:2010zg}.

A lot of progress has been done in understanding the mathematical structure of amplitudes in planar ${\cal N}=4$ SYM.
It is expected that
$n$-point amplitudes are expressed as iterated integrals of differential one-forms~\cite{Chen:1977oja} defined on the space of configurations of points in the three-dimensional projective space $\textrm{Conf}_N(\mathbb{CP}^3)$~\cite{Golden:2013xva,Golden:2014xqa}. 
The simplest instance of iterated integrals that one encounters in amplitudes of ${\cal N}=4$ SYM are the multiple polylogarithms (MPL)~\cite{Goncharov:2001iea,Brown:2009qja} which correspond to iterated integrals over rational functions. It is believed that all maximally helicity violating (MHV) and next-to-MHV amplitudes in planar ${\cal N}=4$ SYM can be expressed in terms of MPLs of uniform transcendental weight~\cite{ArkaniHamed:2012nw}. 

In the Euclidean region, where all Mandelstam invariants are negative, scattering amplitudes in planar ${\cal N}=4$ SYM in MRK
factorise to all orders in perturbation theory into building blocks, which are the impact factor, the Regge trajectory and the central-emission vertex.
The last two describe respectively  the Reggeised gluons in the $t$-channel and the emission of gluons along the $t$-channel ladder formed by the Reggeised gluons. Those building blocks are determined to all orders by the four and five-point amplitudes, and hence scattering amplitudes in MRK are trivial in the Euclidean region~\cite{Brower:2008nm,Bartels:2008ce,DelDuca:2008pj,Bartels:2008sc,Brower:2008ia,DelDuca:2008jg}. In particular, the Regge trajectory is fixed, to all orders in perturbation theory, by the cusp anomalous dimension and the gluon collinear anomalous dimension~\cite{DelDuca:2008pj}.

Starting from six legs, scattering amplitudes exhibit Regge cuts, if the multi-Regge limit is taken after analytic continuation to a specific Minkowski region~\cite{Bartels:2008ce,Bartels:2008sc}. The discontinuity across the cut is described to all orders by a dispersion 
relation closely related to the BFKL  equation, which can be expressed in terms of single-valued functions~\cite{Lipatov:2010ad,Fadin:2011we}. 
In particular, the discontinuity of the six-point amplitude~\cite{Dixon:2012yy} in MRK can be expressed in terms of single-valued harmonic polylogarithms (SVHPL)~\cite{BrownSVHPLs}. More generally, $n$-point amplitudes in MRK are 
described by the moduli space $\mathfrak{M}_{0,n-2}$ of Riemann spheres with $(n-2)$ marked points~\cite{DelDuca:2016lad}. 
The algebra of iterated integrals on $\mathfrak{M}_{0,n}$ factors in such a way that the iterated integrals on $\mathfrak{M}_{0,n}$ can always be expressed in terms of MPLs~\cite{Brown:2009qja} and rational functions with singularities when two marked points coincide. The discontinuity of the $n$-point amplitude in MRK is parametrised by $(n-5)$ conformally invariant ratios.
In the soft limit of any of the transverse momenta of the gluons emitted along the ladder, one of the $(n-5)$
ratios vanishes, and thus two marked points on the Riemann sphere coincide~\cite{DelDuca:2016lad}. 
However, the transverse momenta of the ladder gluons
never vanish. This requirement constrains the iterated integrals that can appear in MRK to single-valued functions, {\it i.e.} linear combinations of products of iterated integrals on $\mathfrak{M}_{0,n}$ (and their complex conjugates) such that all branch cuts cancel, which are single-valued~\cite{BrownSVHPLs,BrownSVMPLs,Brown:2013gia} multiple polylogarithms (SVMPL). The statements 
of Ref.~\cite{DelDuca:2016lad}, which were made at LL accuracy, can be generalised at NLL accuracy and beyond~\cite{DelDuca:2018hrv}.

\section{The Regge limit in ${\cal N}=4$ SYM meets the Regge limit in QCD}

Underpinning the analysis of the $n$-point amplitudes in planar ${\cal N}=4$ SYM in MRK, is Lipatov's
picture~\cite{Lipatov:2009nt} of a BFKL-like dispersion relation, which describes the discontinuity in terms of the Hamiltonian of an integrable open Heisenberg spin chain~\cite{Lipatov:1993yb,Faddeev:1994zg}. Previously, Lipatov had characterised the BFKL equation, and more in general
the exchange of a ladder of $n$ Reggeised gluons in a singlet configuration in QCD at large $N_c$, in terms of the 
Hamiltonian~\cite{Lipatov:1993yb,Lipatov:1993qn,Lipatov:1998as} of a closed spin chain. It should come then as no surprise that,
just like the six-point amplitude of ${\cal N}=4$ SYM in MRK~\cite{Dixon:2012yy}, also the analytic structure of the 
BFKL ladder at LL accuracy in QCD can be described in terms of single-valued iterated integrals on the moduli space ${\cal M}_{0,4}$ of 
Riemann spheres with four marked points, which are SVHPLs~\cite{DelDuca:2013lma}. In this case, the single-valuedness can be traced back to
the fact that neither of the momenta at the ends of the gluon ladder can vanish. That analysis can be extended at NLL accuracy to the BFKL ladder 
in QCD, as well as in full ${\cal N}=4$ SYM~\cite{DelDuca:2017peo}, however it requires a generalisation of the SVHPLs, recently introduced by Schnetz~\cite{Schnetz:2016fhy}. The control of the analytic structure of the BFKL ladder in QCD at NLL accuracy, and the freedom in defining its matter
content, let us prove that there is no gauge theory of uniform and maximal transcendental weight such that 
in momentum space it matches the maximal weight part of QCD, and to identify a set of conditions which allowed us to constrain the field content of theories 
for which the BFKL ladder has maximal weight~\cite{DelDuca:2017peo}.

%
%
%

\end{document}